\documentclass{elsart1p}

\usepackage{graphicx}
\usepackage{amssymb}

\newcommand{\bra}{\langle}
\newcommand{\ket}{\rangle}
\newcommand{\Tr}{\mbox{Tr}\,}

\newcommand{\be}{\begin{equation}}
\newcommand{\ee}{\end{equation}}
\newcommand{\bea}{\begin{eqnarray}}
\newcommand{\eea}{\end{eqnarray}}

\newcommand{\xv}{{\mathbf x}}
\newcommand{\cP}{{\cal P}}
\newcommand{\eps}{\epsilon}

\newcommand{\om}{\omega}
\newcommand{\pv}{{\mathbf p}}

\begin{document}

\begin{frontmatter}

 \title{Two complex problems on the lattice: transport coefficients and 
finite chemical potential\thanksref{label1}}
\thanks[label1]{Invited talk at Strong and Electroweak Matter, 
Amsterdam, the Netherlands, 26-29 August 2008.}
 \author{Gert Aarts}
 \ead{g.aarts@swan.ac.uk}
 \address{Department of Physics, Swansea University, Swansea, United Kingdom}

\begin{abstract}

After a few remarks about the problem of extracting transport coefficients 
from lattice QCD calculations, I report on recent developments in applying 
stochastic quantization and complex Langevin dynamics to field theories 
with a complex action due to a nonzero chemical potential. First results 
demonstrate that the sign problem poses no obstacle for this approach, 
even in the thermodynamic limit. I conclude with a comparison of two 
simple one-link models, describing a euclidean system at finite chemical 
potential and a Minkowski system in real time.

\end{abstract}

\begin{keyword}
 Lattice gauge theory \sep quark-gluon plasma \sep finite-temperature 
field theory 

\PACS 11.15.Ha \sep 12.38.Mh \sep 11.10.Wx
\end{keyword}
\end{frontmatter}

\section{Introduction}
\label{secintro}

The euclidean lattice formulation of QCD provides a nonperturbative 
formulation of the theory of strong interactions, suitable for numerical 
simulations. This allows a detailed study of thermodynamic quantities, 
such as pressure, entropy and susceptibilities, and of the 
finite-temperature crossover between the confined and the deconfined phase 
(see e.g.\ Ref.\ \cite{Schmidt:2008cf}). However, for some problems 
standard methods are not adequate. In this contribution I discuss two of 
these problems. In the first case, the questions asked are not easily 
answered from knowledge of euclidean-time correlation functions alone, but 
instead require access to particular real-time correlators such as 
spectral functions. This is relevant for transport properties of the 
quark-gluon plasma and for in-medium modifications of hadrons. In the 
second case the weight in the euclidean path integral is not real, leading 
to the infamous sign problem. This prohibits the use of importance 
sampling, the framework underlying essentially all methods used to 
generate ensembles in lattice QCD simulations. This is relevant for QCD at 
nonzero baryon density.

\section{Transport and spectral functions}
\label{sectransport}

Conserved currents, such as the energy-momentum tensor $T^{\mu\nu}$ and 
the electromagnetic current $j^\mu$, result in hydrodynamic behaviour on 
long length and timescales. Hydrodynamic structure is best visible in 
current-current correlation functions in real time, such as spectral 
functions. A nonperturbative determination of spectral functions $\rho$ 
using euclidean correlators $G_E$ obtained with lattice simulations, via 
the relation
 \be
 G_E(\tau, \pv) = \int_{0}^{\infty} 
\frac{d\om}{2\pi}\,K(\om,\tau)\rho(\om, \pv),
 \;\;\;\; \;\;\;\;
 K(\om,\tau) = \frac{\cosh[\om(\tau-1/2T)]}{\sinh(\om/2T)},
 \ee
 is quite involved. One widely used approach is the Maximal Entropy Method 
(MEM) \cite{Asakawa:2000tr}. However, in the hydrodynamical regime ($\om, 
|\pv| \ll T$) the problem is particularly difficult due the insensitivity 
of the euclidean correlator to details of the spectral function at small 
$\om$ \cite{Aarts:2002cc}. Moreover, the most commonly used MEM algorithm 
is inherently unstable at small energies and needs to be modified in order 
to be applicable \cite{Aarts:2007wj}. There exist recent lattice 
calculations of the electrical conductivity \cite{Aarts:2007wj}, the shear 
viscosity \cite{Meyer:2007ic} and the bulk viscosity \cite{Meyer:2007dy}. 
Transport and hydrodynamics from the lattice have been discussed in detail 
in two plenary talks at the {\em Lattice} conferences in the past two 
years \cite{Aarts:2007va,Meyer:2008sn}. Therefore I restrict myself to 
some remarks and refer to those contributions for more details.

What euclidean correlators should be considered? 
Kubo relations relate transport coefficients to the slope of spectral 
functions as the energy $\om$ is taken to zero. For example, the shear 
viscosity $\eta$, bulk viscosity $\zeta$, and conductivity $\sigma$ are 
determined by
 \be
 \eta = \lim_{\om\to 0} \frac{\rho^{12,12}(\om)}{2\om},
\;\;\;\;\;\;\;\;\;\;\;\;
 \zeta = \frac{1}{9}\lim_{\om\to 0} \frac{\rho^{ii,jj}(\om)}{2\om},
\;\;\;\;\;\;\;\;\;\;\;\;
 \sigma = \lim_{\om\to 0} \frac{\rho^{11}(\om)}{2\om},
 \ee
in terms of the spectral functions
 \be
 \label{eqrho}
\hspace*{-0.6cm}
 \rho^{\mu\nu,\kappa\sigma}(\om) = \int d^4x\, e^{i\om t} 
\bra [ T^{\mu\nu}(t,\xv),T^{\kappa\sigma}(0)]\ket,
\;\;\;\;
 \rho^{\mu\nu}(\om) = \int d^4x\, e^{i\om t} \bra [ 
j^\mu(t,\xv),j^\nu(0)]\ket. 
 \ee
 In analytical calculations it is sometimes useful to add an exactly 
conserved charge, such as ${\cal E} = \int d^3x\, T^{00}$ or $Q=\int 
d^3x\, j^0$, to the operators in Eq.\ (\ref{eqrho}). For instance, for the 
bulk viscosity one may consider correlators of $T^\mu_\mu$ instead of 
$T^{ii}$ \cite{Meyer:2007dy,Moore:2008ws}. While this is harmless in 
analytical calculations, it makes a difference when reconstructing 
spectral functions from the lattice. The reason is that lattice 
correlators and spectral functions are affected in a particular manner: if 
a conserved charge is added, lattice correlators will receive a 
$\tau$-independent contribution of the form $T\Xi$, where $\Xi$ is the 
corresponding charge susceptibility, while spectral functions will receive 
a contribution $\Xi 2\pi\om\delta(\om)$. This singular contribution at the 
origin violates the smoothness condition assumed by most methods used to 
obtain spectral functions from euclidean correlators and should therefore 
be avoided.
 This issue was discussed already some time ago in the context of meson 
spectral functions and the current $j^\mu = \bar\psi \gamma^\mu \psi$ 
\cite{Karsch:2003wy,Aarts:2005hg}. For massless noninteracting fermions 
the delta function contribution at the origin is equal for 
$\rho^{00}(\om)$ and $\rho^{ii}(\om)$ and cancels in the combination 
$\rho^\mu_\mu(\om)$. However, interactions will smear out this delta 
function in the case of $\rho^{ii}$ (resulting in a finite conductivity), 
whereas $\rho^{00}$ is unchanged due to charge conservation (only the 
value of the susceptibility is affected). The cancelation therefore no 
longer holds and $\rho^\mu_\mu(\om)$ is not smooth as $\om\to 0$.

As a second remark, let me note again that euclidean correlators are 
insensitive to details of spectral functions when the transport 
contribution is narrow and most of the weight is concentrated at $\om \ll 
T$. It is not so difficult to find cases where access to transport 
coefficients from euclidean correlators is virtually impossible. This has 
been discussed for weakly coupled theories \cite{Aarts:2002cc}, heavy 
quark diffusion \cite{Petreczky:2005nh} and, more recently, bulk viscosity 
in a weakly coupled theory or near a second order phase transition 
\cite{Moore:2008ws}. (For a lattice study of energy-momentum tensor 
correlators near the second order transition in SU(2) gauge theory, see 
Ref.\ \cite{Huebner:2008as}.) From the viewpoint of the lattice, it would 
be more interesting to find instead examples where the transport peak is 
not narrow such that transport coefficients might potentially be 
accessible. So far this has been addressed mostly in theories with a 
gravity dual in the strong coupling limit \cite{Teaney:2006nc, 
Kovtun:2006pf, Myers:2007we}. Since these results provide an important 
motivation, more examples would be welcome.

\section{Finite chemical potential and the sign problem}
\label{secsign}

I now turn to euclidean field theories with a complex action due to the 
presence of a chemical potential. In this case the weight in the path 
integral is not real and importance sampling cannot be used. This is 
commonly referred to as the sign problem.
If the weight is written as $e^{-S} = |e^{-S}|e^{i\varphi}$, we may 
consider the partition functions of the full and the phase-quenched (pq)
theories, defined as
 \be
 Z_{\rm full} = \int D\phi\, e^{-S}, 
 \;\;\;\; \;\;\;\; \;\;\;\;
 Z_{\rm pq} = \int D\phi\, |e^{-S}|.
 \ee
 The phase-quenched theory has a real positive weight and is accessible 
with standard techniques. The sign problem is best demonstrated by the 
expression for the expectation value of the phase factor in the 
phase-quenched theory,
 \be
 \label{eq5}
 \bra e^{i\varphi}\ket_{\rm pq} = \frac{ Z_{\rm full}}{Z_{\rm pq}} = 
e^{-\Omega \Delta f},
\ee
 where $\Omega$ is the (four)volume and $\Delta f$ is the difference 
between the free energy densities in the full and the phase-quenched 
theories. Eq.\ (\ref{eq5}) shows that the average phase factor goes to 
zero in the thermodynamic limit, resulting in an overlap problem between 
the full and the phase-quenched theories. In other words, the important 
configurations in the full theory differ in an essential way from those in 
the phase-quenched case. The question is how to find those relevant 
configurations in a numerical simulation.

Recently, this problem was reconsidered \cite{Aarts:2008rr,Aarts:2008wh} 
using stochastic quantization and complex Langevin dynamics 
\cite{Parisi:1980ys,Parisi:1984cs,Klauder:1985a}. This approach has a long 
history and was studied intensely in the 80's \cite{Damgaard:1987rr}. It 
was revived in the context of nonequilibrium quantum field dynamics in 
Minkowski spacetime in Refs.\ 
\cite{Berges:2005yt,Berges:2006xc,Berges:2007nr} and subsequently applied 
to euclidean theories at nonzero chemical potential. We have considered 
SU(3) gauge theory with heavy quarks \cite{Aarts:2008rr} and the scalar 
O(2) model \cite{Aarts:2008wh}, both at nonzero chemical potential.

The essence of this approach is that the weight is obtained as the 
equilibrium distribution of a stochastic process, described by a Langevin 
equation. For a scalar field the Langevin equation reads
 \be
 \phi(x,\theta+\eps) = \phi(x,\theta) -\eps\frac{\delta
S[\phi]}{\delta\phi(x,\theta)} + \sqrt{\eps}\eta(x,\theta),
\ee
while in SU($N$) gauge theories, the links are updated according to 
\be
 U(\theta+\eps) = R(\theta)U(\theta), 
\;\;\;\;\;\;\;\;
 R = \exp\left[i\lambda_a\left(\eps K_a+\sqrt{\eps}\eta_a\right)\right],
\;\;\;\;\;\;\;\;
K_a = -D_aS[U].
\ee
Here $\theta=n\eps$ is the Langevin time, $\eps$ is the Langevin timestep, 
the noise is Gaussian, 
\be \bra \eta(\theta)\ket = 0,
\;\;\;\;\;
\bra \eta(\theta)\eta(\theta')\ket = 2\delta_{nn'},
\ee
 $\lambda_a$ are the Gell-mann matrices and all other indices are 
suppressed. When the action is complex, the dynamics and field variables 
are complexified. Explicitly, a real scalar field is written as $\phi \to 
\phi^{\rm R} + i\phi^{\rm I}$, while an SU($N$) matrix is now an element 
of SL($N$, $\mathbb{C}$). More details can be found in the references 
listed above.

\begin{figure}
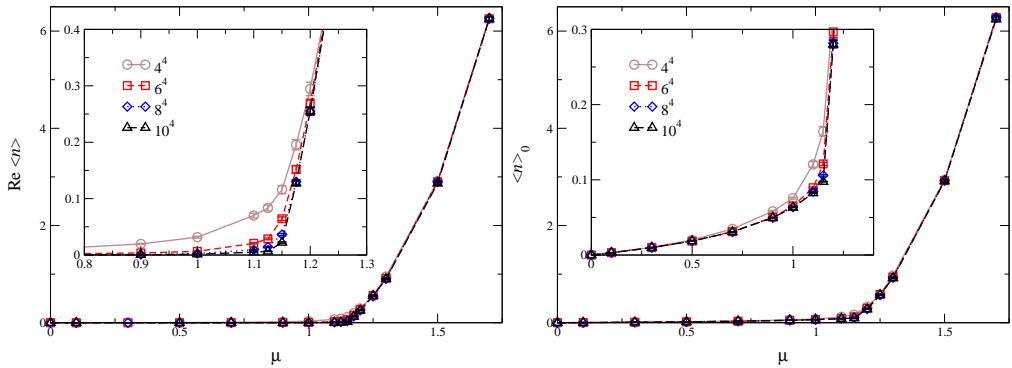

\begin{center}
\includegraphics[height=4.8cm]{fig12.eps}
\includegraphics[height=4.8cm]{fig4.eps}
\end{center}
 \caption{
Real part of the density in the full (left) and the phase-quenched (right) 
relativistic Bose gas as a function of chemical potential on lattices with 
size $N^4$, with $N=4,6,8,10$ ($m=\lambda=1$).
 }
\label{figo2_1}
\end{figure}

I now discuss the results obtained so far. I start with the most recent 
results in the relativistic Bose gas, i.e.\ a complex scalar field with 
O(2) symmetry and a quartic interaction term, at nonzero chemical 
potential \cite{Aarts:2008wh}. The action satisfies $S^*(\mu)=S(-\mu)$. 
Just as QCD, this theory has what is known as a Silver Blaze problem 
\cite{Cohen:2003kd}. At zero temperature the theory is in vacuum as long 
as the chemical potential $\mu$ is below the critical value $\mu_c$ (when 
interactions are ignored, $\mu_c=m$, the minimal energy of excitations). 
When $\mu>\mu_c$, the theory enters a Bose condensed phase and the density 
is nonzero. The strict $\mu$-independence of bulk physical quantities as 
long as $\mu<\mu_c$, even though $\mu$ is present microscopically, is the 
Silver Blaze problem. We have analysed the full and the phase-quenched 
theory using stochastic quantization. The results for the density are 
shown in Fig.\ \ref{figo2_1}. On the left, we show the density in the full 
theory, as a function of the chemical potential, for lattices of size 
$N^4$, with $N=4,6,8,10$. The parameters in the scalar potential are 
$m=\lambda=1$, in lattice units. The inset shows a blowup of the 
transition region. We observe the transition between the groundstate with 
zero density and the Bose condensed phase at $\mu=\mu_c\sim 1.15$. A real 
phase transition can only occur in the thermodynamic limit and we find 
that the transition becomes sharper as the volume is increased, as 
expected. On the right, the density is shown in the phase-quenched theory. 
Here we observe a nonzero density for all values of the chemical 
potential. This is similar to what is expected in phase-quenched QCD (when 
$m_\pi/2<\mu<m_N/3$). The cancelation of the $\mu$-dependence of the 
density in the full theory is due to the phase factor $e^{i\varphi}$ and, 
therefore, the result of the sign problem. We conclude that the complexity 
of the action and the Langevin dynamics deliver precisely what is 
expected.

\begin{figure}
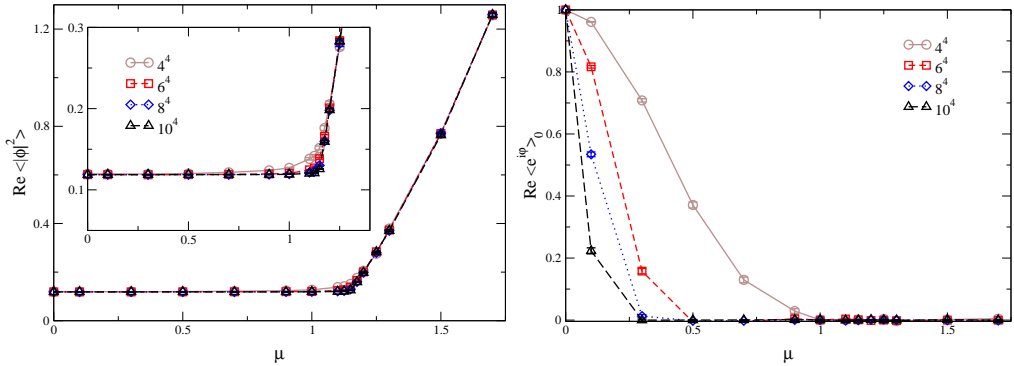

\begin{center}
\includegraphics[height=4.8cm]{fig3.eps}
\includegraphics[height=4.8cm]{fig5.eps}
\end{center}
 \caption{
Real part of the square of the field modulus in the full theory (left) and 
the average phase factor in the phase-quenched theory (right).
 }
\label{figo2_35}
\end{figure}

The $\mu$-independence also hold for other observables. In Fig.\ 
\ref{figo2_35} (left) we show the square of the field modulus $|\phi|^2$ 
as obtained from complex Langevin dynamics. The value at zero chemical 
potential is nonzero due to quantum fluctuations. As $\mu$ is increased, 
the observable remains constant until the critical value is reached. This 
demonstrates the Silver Blaze feature in an impressive manner. On the 
right the average phase factor in the phase-quenched theory is shown. As 
indicated above, this observable gives an indication of the severeness of 
the sign problem: the phase factor goes to zero at larger chemical 
potential on all lattices and vanishes in the thermodynamic limit for all 
nonzero values of the chemical potential. This is precisely how the 
average phase factor is expected to behave \cite{Splittorff:2006fu}. 
Nevertheless, this approach can handle the sign problem without 
difficulties and there is no problem in taking the thermodynamic limit.

Concerning QCD we have considered SU(3) gauge theory coupled to three 
flavours of quarks in the heavy quark limit, starting from Wilson fermions 
in the hopping expansion \cite{Aarts:2008rr}. In this limit, the complex 
fermion determinant is approximated as
 \be
\det M \approx \prod_{\xv}
\det\left( 1 + h e^{\mu/T} \cP_{\xv} \right)^2
\det\left( 1 + h e^{-\mu/T} \cP_{\xv}^{-1} \right)^2,
\ee
 where $h = (2\kappa)^{N_\tau}$ and $\cP_\xv^{(-1)}$ are the (conjugate)
Polyakov loops. We emphasize that the gauge action is preserved 
completely and that the determinant satisfies the basic property $[\det 
M(\mu)]^* = \det M(-\mu)$.

\begin{figure}
\begin{center}
\includegraphics[height=4.8cm]{plot_fsu3_rePPcc_Nt4_ch2.eps}
\includegraphics[height=4.8cm]{plot_fsu3_redensity_Nt4_ch2.eps}
\end{center}
 \caption{
Real part of the (conjugate) Polyakov loops $\bra P\ket$ and $\bra 
P^{-1}\ket$ (left) and the density $\bra
n\ket$ (right) as a function of $\mu$ for QCD in the heavy quark limit, 
with $\beta=5.6$, $\kappa=0.12$, and $N_f=3$ on a $4^4$ lattice.
 }
\label{figfsu3}
\vspace*{0.5cm}
\begin{center}
\includegraphics[height=4.7cm]{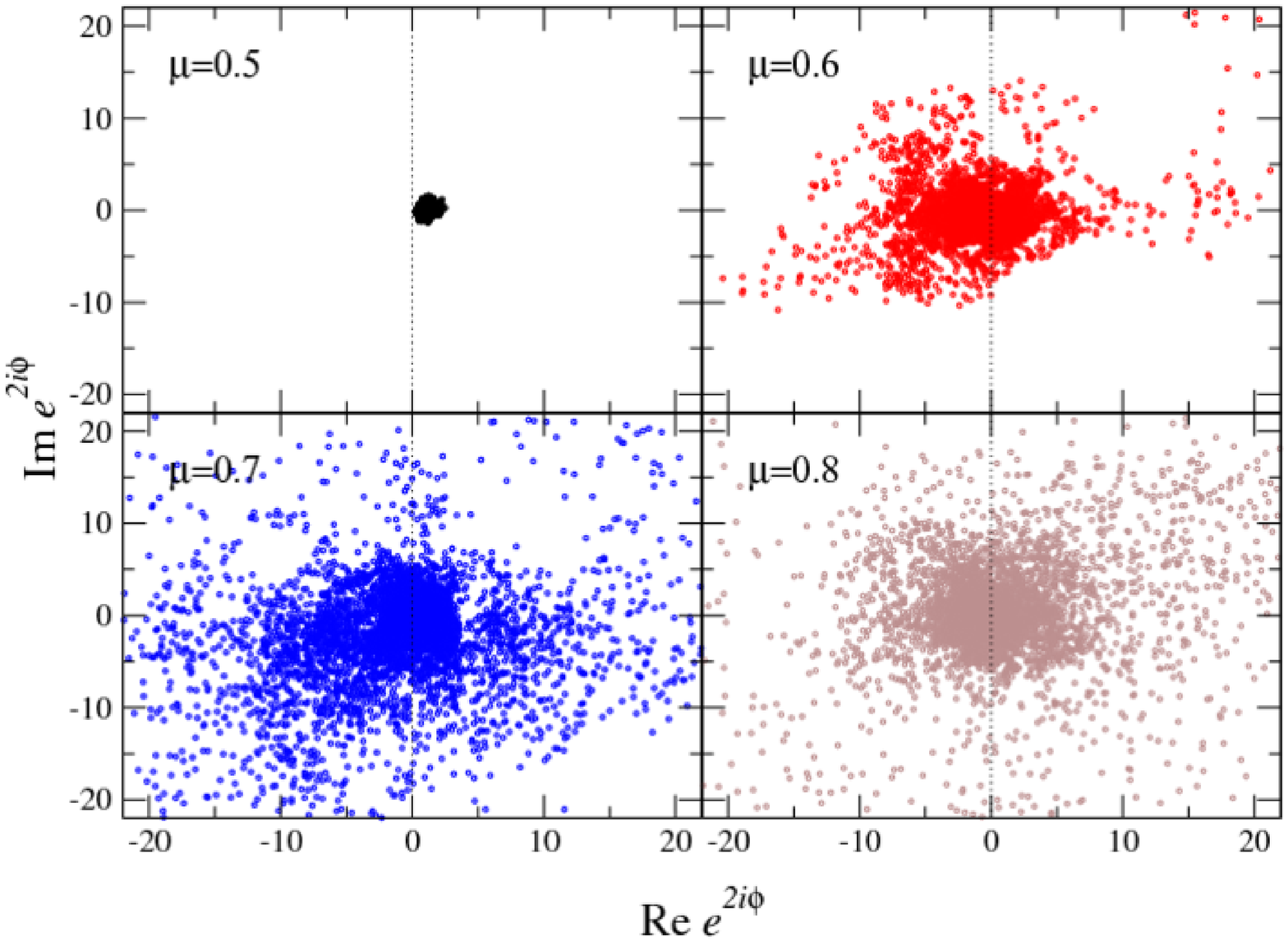}
\includegraphics[height=4.7cm]{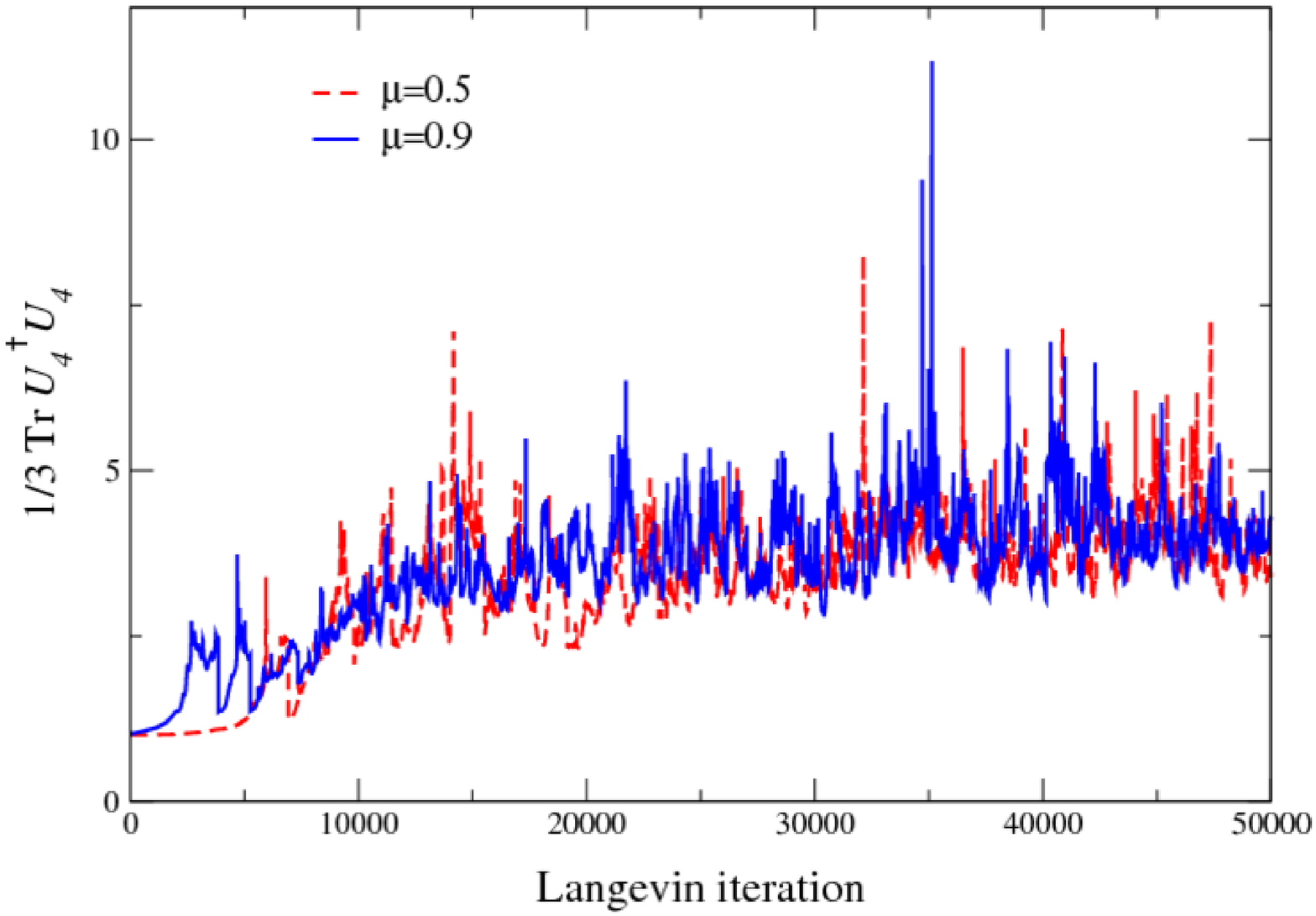}
\end{center}
 \caption{
Left: Scatter plot of $e^{2i\phi} = \det M(\mu)/\det
M(-\mu)$ during the Langevin evolution for various values of $\mu$.
Right:
Deviation from SU(3), $\Tr U_4^\dagger U_4/3$ during the Langevin
evolution, for $\mu=0.5$ and $0.9$.
 }
\label{figscat}
\end{figure}

First results in this theory are obtained on a lattice of size $4^4$, with 
fixed $\beta=5.6$ and $\kappa=0.12$. In Fig.\ \ref{figfsu3} we show the 
(conjugate) Polyakov loops (left) and the density (right) as a function of 
chemical potential. The results indicate a transition from a confined 
low-density phase to a deconfined high-density phase. Again we address the 
severeness of the sign problem by studying the phase of the determinant. 
In this case we present the phase factor $e^{2i\phi} = \det M(\mu)/\det 
M(-\mu)$ in the full theory, and not in the phase-quenched theory. A 
scatter plot of the real and imaginary parts of this observable is shown 
in Fig.\ \ref{figscat} (left). When the chemical potential is zero, the 
phase factor is equal to 1. At larger chemical potential, we find that 
phase fluctuations increase dramatically. Yet, observables such as the 
Polyakov loop and the density are under control, with reasonable errors. 
During the complex Langevin evolution, link variables are no longer in 
SU(3), instead they take values in SL(3, $\mathbb{C})$. This can analysed 
by computing $\Tr U_4^\dagger U_4/3$, which equals 1 in SU(3) and is $\geq 
1$ in SL(3, $\mathbb{C})$. We show this quantity as a function of Langevin 
time in Fig.\ \ref{figscat} (right). We observe that after the initial 
thermalization stage it is clearly distinct from 1, but remains bounded. 
This indicates that the sequence of configurations generated with complex 
Langevin dynamics differs in an essential way from configurations 
constrained to be in SU(3). For more discussion, see Refs.\ 
\cite{Aarts:2008rr,Aarts:2008tc}.

\section{Chemical potential versus Minkowski dynamics}

\begin{figure}[t]
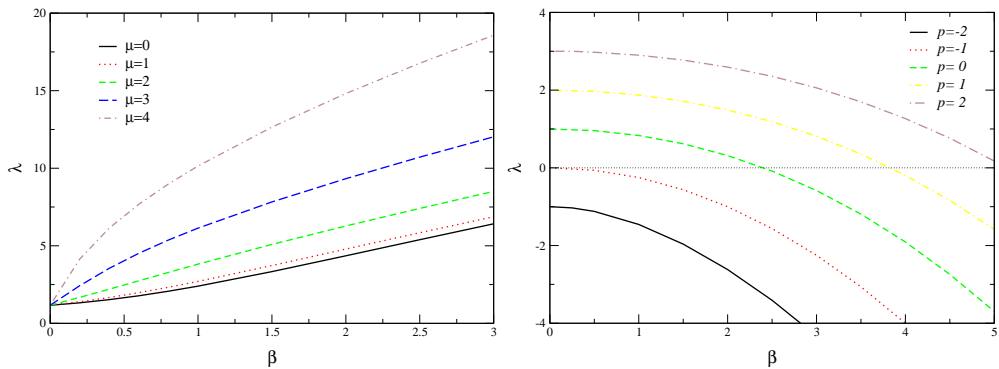

\begin{center}
\includegraphics[height=4.8cm]{fokker_eigenvalues_k0.5_mu0-4.eps}
\includegraphics[height=4.8cm]{fokker_cmplxbeta_p.eps}
\end{center}
 \caption{Smallest nonzero eigenvalue of the complex Fokker-Planck 
operator in the one-link U(1) model as a function of $\beta$ for various 
values of $\mu$ at $\kappa=1/2$ in the euclidean case (left) and for 
various values of the reweighting parameter $p$ in the Minkowski case 
(right). 
 }
 \label{fig:fp}
\end{figure}

At zero chemical potential the euclidean action is real. In this case one 
can use real Langevin evolution and apply standard proofs to demonstrate 
that the method will converge to the correct distribution $e^{-S}$. Zero 
chemical potential provides therefore a useful reference point, see e.g. 
Fig.\ \ref{figo2_35} (left). In the case of Minkowski (real-time) 
dynamics, where the weight in the path integral is $e^{iS}$, such a 
reference point is absent. Heuristically, one may therefore expect real 
time to be more difficult than nonzero chemical potential. To quantify 
this, we compared two simple one-link U(1) models \cite{Aarts:2008rr}, 
building on the work in Ref.\ \cite{Berges:2007nr}. The partition 
functions are
 \be
 Z =  \int_{-\pi}^\pi\frac{dx}{2\pi}\, e^{\beta\cos x}
\left[   1+\kappa \cos(x-i\mu) \right],
\ee
 at nonzero $\mu$ and
\be
 Z =  \int_{-\pi}^\pi\frac{dx}{2\pi}\, e^{i\beta\cos x + ipx},
\ee
 in real time. The term $px$, with $p$ integer, is a reweighting term, 
used to stabilize the Langevin dynamics \cite{Berges:2007nr}. In the 
euclidean case the Fokker-Planck equation for the complex distribution 
$P(x,\theta)$ reads
 \be
 \frac{\partial}{\partial \theta} P(x,\theta) = 
 \frac{\partial}{\partial x} \left(
 \frac{\partial}{\partial x} + \frac{\partial S}{\partial x} 
 \right) P(x,\theta).
\ee
 The zero eigenvalue, corresponding to the stationary distribution 
$e^{-S}$, always exists. The solution of the Fokker-Planck equation can 
therefore be written as
\be
 P(x,\theta) = \frac{e^{-S}}{Z} + \sum_{\lambda\neq 0} 
e^{-\lambda\theta}P_\lambda(x).
\ee
 For the Minkowski case, $-S$ is replaced with $iS$ in the equations 
above. 
Convergence of the complex Fokker-Planck equation is determined by the 
eigenvalues $\lambda$, which we studied numerically \cite{Aarts:2008rr}.
 The smallest nonzero eigenvalue is shown in Fig.\ \ref{fig:fp} as a 
function of $\beta$ in the euclidean case for various values of $\mu$ 
(left) and the Minkowski case for various values of $p$ (right). In the 
first case, one sees that the eigenvalues are strictly positive, for all 
values of $\beta$ and $\mu$, indicating convergence of the complex 
Fokker-Planck equation. In the second case, eigenvalues turn negative for 
larger values of $\beta$; the value of $\beta$ where this occurs depends 
on the parameter $p$. While these results do not prove or disprove 
convergence for the complexified Langevin dynamics (for this the real 
Fokker-Planck equation for the real distribution $\rho(x,y,\theta)$ has to 
be studied), they are an indication of the difference between the two 
theories when applying complex Langevin dynamics.

\section{Acknowledgments}

It is a pleasure to thank the organizers for the stimulating Conference. 
In particular I am grateful to Jan Smit, not only for this event, but also 
for the continuous support since 1995. I thank Nucu Stamatescu for 
collaboration and encouragement and Simon Hands, Harvey Meyer and Erhard 
Seiler for discussion. This work is supported by STFC.


\end{document}